\documentclass[letterpaper, 10 pt, conference]{IEEEtran}
\newcommand{\calA}{\mathcal{A}}

\newcommand{\calZ}{\mathcal{Z}}

\newcommand{\calN}{\mathcal{N}}

\newcommand{\pr}{\mathbb{P}}

\newcommand{\one}{\mathbbm{1}}

\usepackage{mathrsfs}
\usepackage{amssymb}
\usepackage{amsmath,bm}

\usepackage{mathrsfs}
\usepackage{graphicx}
\usepackage{enumerate}
\usepackage{eurosym}
\usepackage{amssymb}
\usepackage{amsmath,bbm}
\usepackage{amsfonts}
\usepackage{epstopdf}
\usepackage{epsf,subfigure}
\usepackage{psfrag}
\usepackage{graphics}
\usepackage{color} 
\usepackage{cite}
\usepackage{array,multirow,pbox}
\usepackage{enumitem}

\newtheorem{remark}{Remark}

\graphicspath{{../figures/}}

\begin{document}
%

\title{\vspace{0.5in}Occupancy-Driven Stochastic Decision Framework for Ranking Commercial Building Loads}

\author{\IEEEauthorblockN{Milan Jain, Soumya Kundu, Arnab Bhattacharya, Sen Huang, Vikas Chandan, Nikitha Radhakrishnan,\\ Veronica Adetola and Draguna Vrabie}
\IEEEauthorblockA{Pacific Northwest National Laboratory, Richland, WA, USA\\
Email: \{milan.jain,\,soumya.kundu,\,arnab.bhattacharya,\,sen.huang,\,vikas.chandan,\,nikitha.radhakrishnan\}@pnnl.gov,\\ veronica.adetola@pnnl.gov, draguna.vrabie@pnnl.gov}}


\maketitle

\begin{abstract}
For effective integration of building operations into the evolving demand response programs of the power grid, real-time decisions concerning the use of building appliances for grid services must excel on multiple criteria, ranging from the added value to occupants' comfort to the quality of the grid services. In this paper, we present a data-driven decision-support framework to dynamically rank load control alternatives in a commercial building, addressing the needs of multiple decision criteria (e.g. occupant comfort, grid service quality) under uncertainties in occupancy patterns. We adopt a stochastic multi-criteria decision algorithm recently applied to prioritize residential on/off loads, and extend it to i) complex load control decisions (e.g. dimming of lights, changing zone temperature set-points) in a commercial building; and ii) systematic integration of zonal occupancy patterns to accurately identify short-term grid service opportunities. We evaluate the performance of the framework for curtailment of air-conditioning, lighting, and plug-loads in a multi-zone office building for a range of design choices. With the help of a prototype system that integrates an interactive \textit{Data Analytics and Visualization} frontend, we demonstrate a way for the building operators to monitor the flexibility in energy consumption and to develop trust in the decision recommendations by interpreting the rationale behind the ranking.
\end{abstract}


%
\IEEEpeerreviewmaketitle


\section{Introduction}
With innovation in smart cities on the rise, the integration of building operation with the operation of the modern power grid through its various demand flexibility programs is inevitable. According to some estimate, buildings contribute to 75\% of total electricity demand and up to 80\% of the peak demand in the USA \cite{conti2018annual}. Therefore, it is critical to integrate their operations with that of the power grid to make the connected building-grid infrastructure more energy-efficient and resilient, leading to the concept of \textit{grid-interactive efficient buildings} \cite{roth2019grid}. Power grid utilities are increasingly recognizing flexible and efficient buildings as one of the lowest-cost energy resources, with investments in the range of billions of US dollar per year towards developing energy efficiency programs in smart buildings and communities \cite{neukomm2019grid}. However, since one of the main purposes of a building is to create a comfortable and convenient environment for its occupants, the problem of integrating buildings into the grids is not straightforward. The task of identifying and engaging building electrical appliances and control options for additional energy flexibility creates a multi-criteria decision problem for the building operators (or, owners), striving to balance the occupant needs with demand response requests. 
%
%
%
%
%
Related prior research has primarily focused on prioritization of residential loads for demand response, with relatively simplistic control options such as switching on/off appliances. Several of these works consider the problem from the perspective of an aggregator of residential switching appliances (e.g. air-conditioners, water-heaters), and propose different prioritized selection schemes \cite{hao2015aggregate,vivekanathan2015,espinosa2017aggregate,nandanoori2018prioritized,hu2020priority}. Other efforts such as \cite{jin2017,kundu2021stochastic} have looked into prioritization of heterogeneous collection of switching appliances within a single residential household (including thermostatic loads, electric vehicles, and plug-loads), while \cite{azar2015aggregated} considered a flexible framework with a collection of one or more residential households.

In comparison, the problem of prioritized selection of load control alternatives in commercial buildings has received much less attention. This is largely due to the fact that the problem is particularly challenging for commercial buildings which are often characterized by multiple zones and shared usage of resources; uncertain and dynamic occupancy patterns in each zone; as well as diversity in appliance control options which go beyond simplistic on/off commands. Nevertheless, some existing works such as \cite{weng2011managing,nutaro2016simulation,kim2016behind} have considered prioritized selection of commercial building loads. The authors in \cite{weng2011managing} proposed a rule-based bi-level (day-time and night-time) priority scheme for plug-loads, incorporating advanced sensing capabilities provided by \textit{smart energy meters} and \textit{passive infrared sensors}. A scheduling algorithm for selection of rooftop air-conditioning units based on their energy requests was proposed in \cite{nutaro2016simulation} to limit peak power demand in small and medium commercial buildings. Similar problem of curtailing rooftop units in small/medium commercial buildings was also considered in \cite{kim2016behind}, in which the authors proposed a heuristic scheme based on \textit{analytic hierarchy process}. The proposed scheme, however, relies heavily on user-supplied (and pre-defined) pair-wise relative weights  on various ranking criteria, and is therefore prone to producing inconsistent results whenever the pairwise comparisons induce flawed logical inferences. 
Majority of the related work in the space of commercial buildings appears to rely on rule-based and/or heuristics, apply to specific (and simplistic) load control options, and disregard the uncertainty and variability of the occupancy patterns which primarily drive energy usage in commercial buildings.

\begin{figure}
    \centering
    \includegraphics[width=\columnwidth]{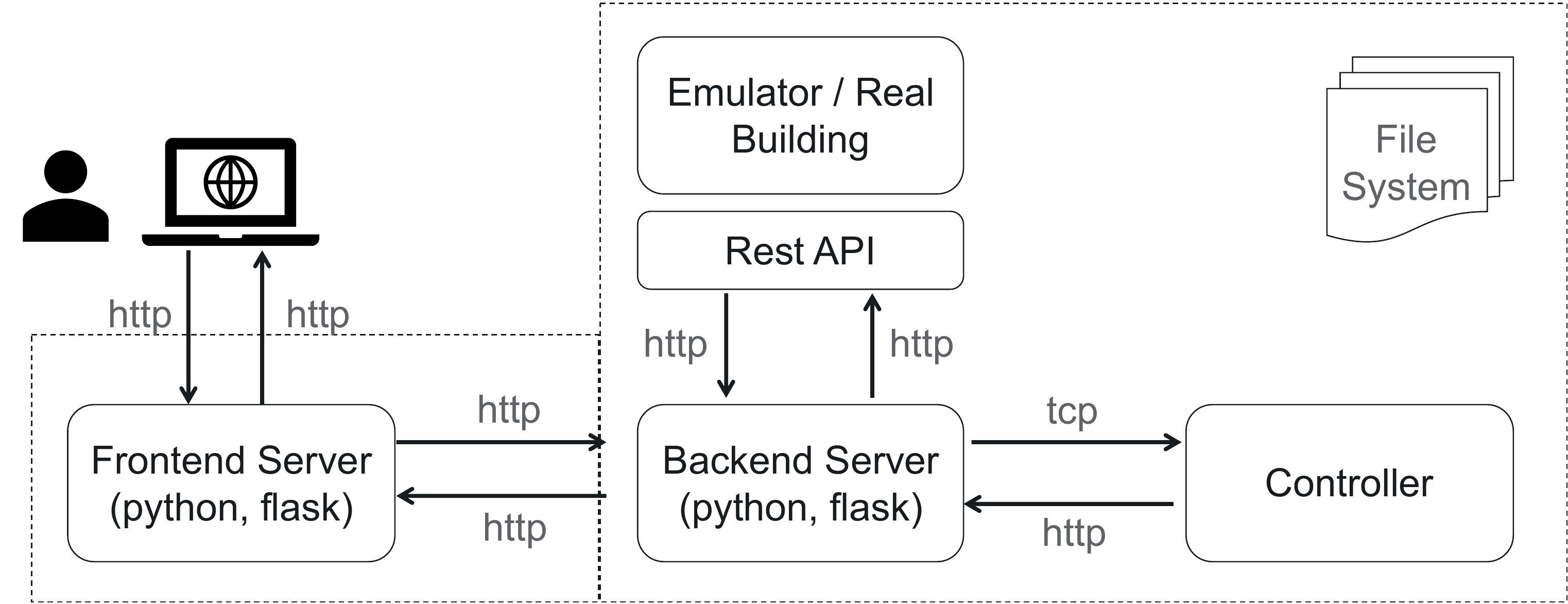}
    \caption{
    System integration: the software prototype used for load prioritization and control testing includes four major components - a frontend, a backend server, a controller and a building emulator.}
    \label{fig:system_architecture}
\vspace{-0.2in}
\end{figure}

In this paper, we present a data-driven stochastic decision-support framework to dynamically rank load control alternatives in a commercial building, addressing the needs of multiple decision criteria (e.g. occupant comfort, grid service quality) under uncertainties in occupancy patterns. The proposed framework is an extension of a stochastic multi-criteria decision-making (MCDM) algorithm which has so far been applied applied to prioritizing residential on/off loads \cite{vivekanathan2015,kundu2021stochastic}. In particular, we extend the MCDM-based dynamic load prioritization algorithm to: i) consider complex control decisions (e.g. dimming of lights, changing zone temperature set-points) in a commercial building; and ii) systematically integrate zonal occupancy patterns to better identify short-term (and time-varying) opportunities for grid service participation. 
The proposed stochastic decision-support and control testing system (Fig.\,\ref{fig:system_architecture}) has four major components: a \textit{frontend}, a \textit{backend}, a \textit{controller} and an \textit{emulator}. The \textit{frontend} hosts a \textit{Data Analytics and Visualization} tool which serves as an interactive user-interface allowing building operators and occupants to enter choices, visualize decision recommendations, and interpret the rationale behind those recommendations. The \textit{backend} server manages the control and data-flow between different modules. The \textit{controller} executes the data analytics engine for offline and online modeling and ranking algorithms, and computes the control decisions; while the \textit{emulator} serves as a virtual testbed performing dynamic simulations of commercial building models. The performance of the proposed ranking and recommendation methodology is demonstrated on demand response scenarios of curtailment of heating, ventilation and air-conditioning (HVAC), lighting and plug-loads in a multi-zone commercial building. 
The rest of the paper is organized as follows. Sec. \ref{sec:problem} describes the problem setup, while major components of the data-driven prioritization algorithm are presented in Sec.\,\ref{sec:load_prioritization}. A description of the simulation framework for system prototyping and performance testing is provided in Sec.\,\ref{sec:system_architecture}.  Sec.\,\ref{sec:results} presents numerical simulation results in a commercial building scenario. Sec.\,\ref{sec:conclusion} concludes the article.

\section{Problem Setup}
\label{sec:problem}

Suppose there are $N$ different control knobs available to the building operator, with each control knob $i$ offering $D_i$ number of (discretized) control settings. For example, a control knob could be the lighting-level in a zone (offering various dimming options), the zonal temperature set-point (allowing discrete changes), or the on/off command to a plug-load (with binary control options). Together, the available control alternatives offered by all the control knobs can be denoted by the set: 
\begin{align*}
\calN\!:=\!\left\lbrace (i,d_{ij})\,\left|i\!\in\!\lbrace 1,\dots,N\rbrace, d_{ij}\!\in\!\lbrace 1,\dots,D_i\rbrace\,\forall i\right.\right\rbrace
\end{align*}
where each tuple $(i,d_{ij})$ represents a control knob and an associated control setting. We denote the number of all control alternatives as $N\!=\!\sum_i D_i$\,. Note that this abstraction allows for generalized (and complex) control commands available in a commercial building, as opposed to the conventional approach of considering only on/off type controls.

The main goal of the stochastic ranking problem is to balance the needs of different stakeholders, such as the desired comfort and convenience of the building occupants, and the contractual bindings and/or incentives of the demand response services for the building owners/operators. Suppose there are $A$ different ranking criteria which form the set $\calA\!:=\!\lbrace 1,\dots,A\rbrace$. Each control alternative is evaluated based on its impact on the buildings performance along those different criteria. Considering load curtailment as the demand response service in this paper, we select two ranking criteria: i) \textit{`comfort'}: quality of end-user experience (e.g. room temperature) resulting from the execution of the control alternative; and ii) \textit{`curtailment'}: the reduction in building load effected by the execution of the control alternative. In the following, we describe how to assign quantitative scores to each control alternative based on its performance along the above-mentioned (qualitative) criteria. Each criteria score is normalized from 0 to 1, with higher scores reflecting better performance.

\begin{figure}[thpb]
\centering
\includegraphics[width=0.8\linewidth]{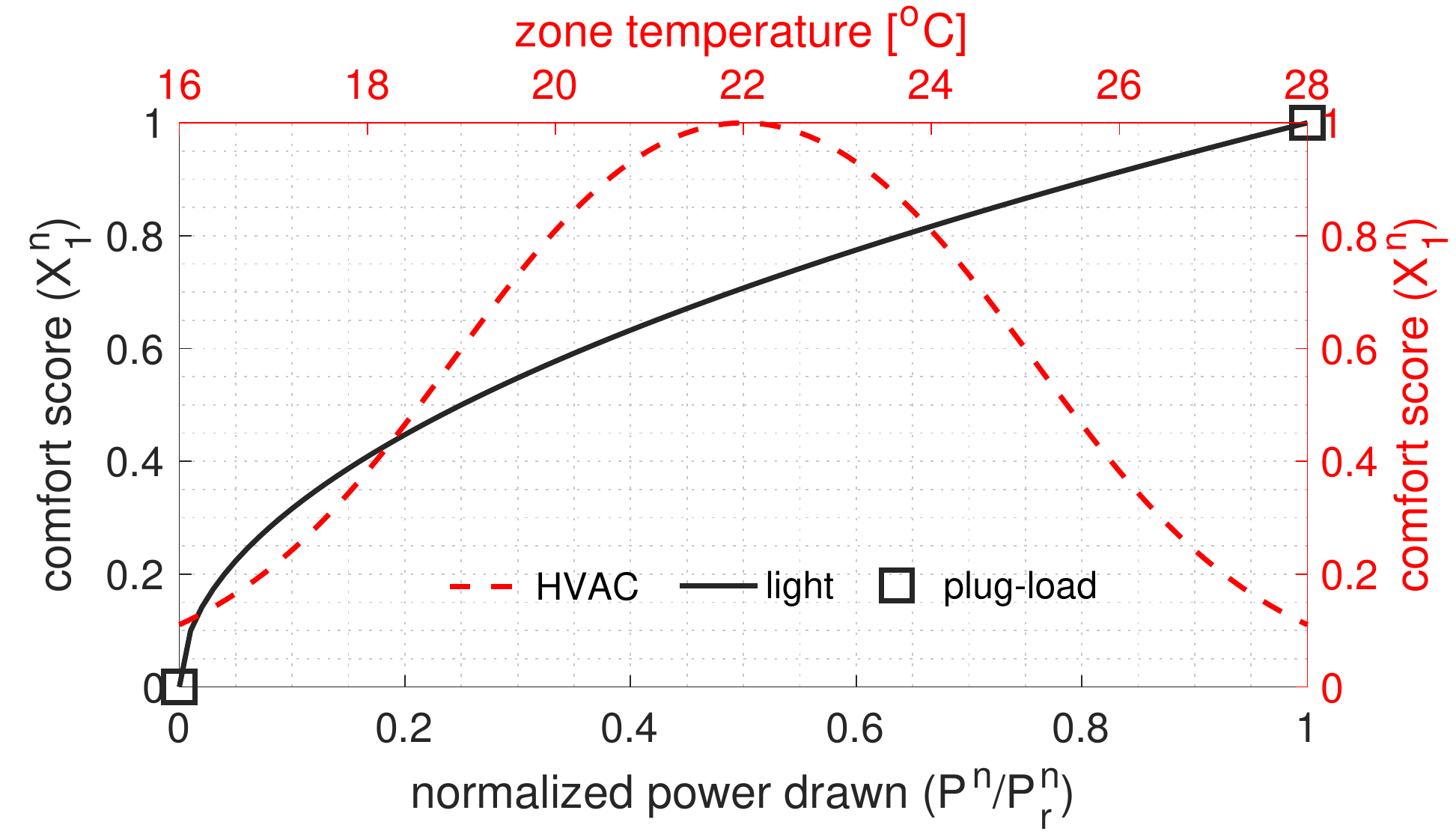}
\caption{Illustration of the perceived comfort scores for three different types of loads - HVAC, dimmable lights, and plug-loads, as given by \eqref{E:score_comfort_ac}-\eqref{E:score_comfort_plug}.}
\label{fig:comfort}
\end{figure}

\textbf{Comfort Scores, $X_1^n$:} The comfort score of a control alternative $n\!\in\!\calN$ quantifies its impact on the occupants' convenience, with higher values of $X_1^n$ representing better comfort. Below we explain the adopted comfort scoring methodology for HVAC, lighting and plug-loads, with an illustrative example of each of the three cases shown in Fig.\,\ref{fig:comfort}.
\begin{enumerate}[leftmargin=0.2in,label=\emph{\alph*)}]
    \item HVAC: Following the findings of the occupants' survey \cite{jin2017}, the thermal comfort associated with every zonal temperature set-point control alternative $n\!\in\!\calN$ is modeled as a function of the difference between the zone temperature ($T^n[k]$, with $k$ as the discrete time-instant) and a desired (`most comfortable') temperature for the zone ($T_s^n$): 
    \begin{align}\label{E:score_comfort_ac}
    X_1^n[k] &=\frac{\alpha_1^n+1/\alpha_1^n+2}{\alpha_1^n+1/\alpha_1^n+\left(\alpha_1^n\right)^{\Delta T^n[k]}+\left(\alpha_1^n\right)^{-\Delta T^n[k]}}\,,\notag\\
        \text{where}~&~\Delta T^n[k]:=\left(\,T^n[k]-T_s^n\right)/\delta T^n\,,
        \end{align}
    $\alpha_1^n\!>\!1$, and $\delta T^n$ models the degradation of perceived comfort at higher and lower temperatures relative to the `most comfortable' temperature. As an illustrative example, Fig.\,\ref{fig:comfort} shows the thermal comfort scores using the following parameter values: $\left(\alpha_1^n,\,T_s^n,\,\delta T^n\right)=\left(10,\,22^oC,\,3^oC\right)$\,.
    
    
    \item Dimmable Lights: The utility of the lighting loads is related to the illuminance of the concerned zone, which in turn is a function of the luminous flux of the lighting load. According to \cite{rea2000iesna}, the perceived brightness by an observant is a square root of the measured (actual) brightness. Since the measured brightness of a dimmable light varies roughly linear with its power consumption, we propose to quantify the comfort score related to dimmable lights as follows:
    \begin{align}\label{E:score_comfort_light}
        X_1^n[k] &= \sqrt{P^n[k]/P_r^n}
    \end{align}
    where $P_r^n$ and $P^n[k]$ denote, respectively, the rated and the actual (with dimming) power drawn by the light.
    
    \item Plug-Loads: For the plug-loads (such as computers), we assume a relatively straightforward comfort scoring methods. Since the plug-loads typically offer only on/off control commands, and are mostly in use whenever the zone is occupied (e.g. computers in an office room), we assign:
    \begin{align}\label{E:score_comfort_plug}
        X_1^n[k] &= 1-P^n[k]/P_r^n, \text{ where } P^n[k]\in \lbrace 0,P_r^n\rbrace
    \end{align}
    
\end{enumerate} 

Note that the comfort scores as defined above are impacted by the choice of control alternatives for the respective loads, e.g. the power consumption for plug-loads and dimmable lights, the temperature set-point for the HVAC. 

\begin{remark}[\textbf{Role of Occupancy Forecast}] The comfort scoring methods proposed above only apply when the zone is occupied. When a zone is unoccupied, the corresponding comfort scores default to unity (i.e. maximum), representing a trivial opportunity for load curtailment. A predictive model of zonal occupancy can, therefore, help accurately identify short-term load curtailment opportunities. More on this in Sec.\,\ref{sec:load_prioritization}.
%
\end{remark}
    
\textbf{Curtailment Scores, $X_2^n$:} The curtailment score of a control alternative $n\!\in\!\calN$ quantifies its impact on reducing the commercial building load. The reduction in building load due to a control action is trivially estimated for the lighting and plug-loads, based on the knowledge of their rated (maximum) power. For HVAC, however, the relation between the zonal set-point changes to the building load is not immediate, and needs to be estimated from experimental/historical data. Sec.\,\ref{sec:load_prioritization} presents a method to learn the sensitivities of the HVAC load to changes zonal temperature set-point changes. For convenience, we denote by $P^n_r$ the estimated load reduction due to the execution of the control alternative $n$. While curtailment score can be modeled as any increasing function of $P_r^n$, we use the following quantitative definition:
    \begin{align}
        X_2^n&=\exp{\left(-\log\left(\alpha_2^n\right) {\log\left(\overline{P}/P_r^n\right)} / {\log\left(\overline{P}/\underline{P}\right)}\right)},
    \end{align}
    where $\alpha_2^n\!>\!1$; and $\overline{P}$ and $\underline{P}$ are, respectively, the largest and the smallest of the $p_r^n$ values. Note that, by its definition, the curtailment score $X_2^n$ takes values from $1/\alpha_2^n$ to 1. The use of logarithm brings some parity between alternatives whose power reduction ratings vary in order(s) of magnitude (e.g. HVAC and personal computers). For similar load control alternatives, however, simple affine relations can be used.

\textbf{Score Prediction and Ranking Alternatives}: Based on the quantitative measures of the performance scores described above, coupled with historical and/or experimental datases of time-series measurements from available sensors (e.g. temperature, power consumption, occupancy), it is possible to develop predictive models the impacts of the various control choices on the different criteria (occupancy comfort and curtailment amount). These predictive models of performance scores are then used by the stochastic ranking algorithm to recommend control options, as described in Sec.\,\ref{sec:load_prioritization}.
\section{Multi-Criteria Stochastic Prioritization}
\label{sec:load_prioritization}


The proposed prioritization algorithm has two major components: a \textit{learning} component, and a \textit{ranking} component. 

\subsection{Data-Driven Learning of Predictive Models}\label{sec:modeling}

In the learning phase, typically offline (but can also be run online in an adaptive fashion), probabilistic predictive models of the performance scores due to the control implementations are identified from historical (and/or experimental) time-series data-sets representing various sensor measurements, contextual information (e.g. weather conditions, weekday/weekends), and occupant feedback-based performance scores (e.g. comfort). There are two parts to this learning problem: identifying static parameters (such as appliance power ratings), and identifying dynamic models (such as time-varying occupancy patterns).

\begin{figure}[thpb]
\centering
\includegraphics[width=0.9\linewidth]{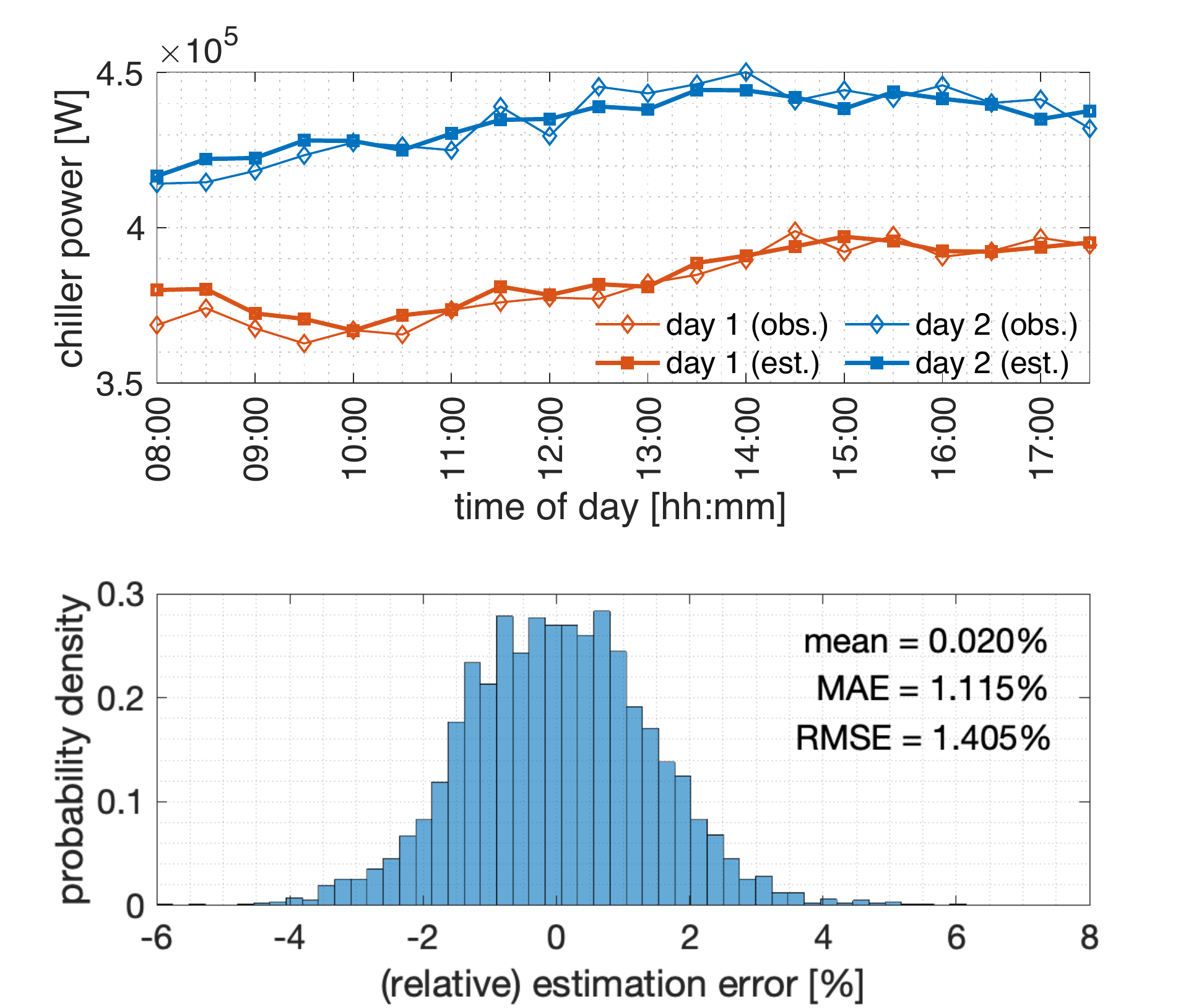}
\caption{(Chiller power estimation) Top plot shows examples of observed and estimated chiller power for two different days. Bottom plot shows the probability densities of the estimation error (relative to the observed power).}
\label{fig:chiller}
\end{figure}

\textbf{Static Parameter Identification:} The power ratings and the allowable levels of power consumption for the various (dimmable) lighting and plug-loads can be identified directly from the historical time-series measurements. HVAC load, however, presents a complicated scenario since the power consumption is measured at the whole building level, while the control options (such as the temperature set-points) reside at the zone levels. Therefore a system identification is required to map the changes in zonal temperature set-points to the changes in the HVAC power consumption. Due to the complexity of HVAC operations in a multi-zone commercial building, such a learning task is often non-trivial. In this paper, however, it suffices for us to focus only on (very) hot weather conditions such that chiller is the predominant HVAC components contributing to its power consumption, and assume the following affine relation between the chiller power ($P_\text{chiller}$), the temperature set-points ($T_{\text{set},z}$) for every zone-$z$, and the outdoor air temperature $T_\text{out}$:
\begin{align}\label{E:chiller}
    P_\text{chiller}[k]\approx \beta_0+\beta_\text{out}\,T_\text{out}[k]+\sum_{z\in\calZ}\beta_z\,T_{\text{set},z}[k]
\end{align}
where $\calZ$ is a set of all the zones in the building, and $\beta_z\!=\!\partial P_\text{chiller}/\partial T_{\text{set},z}$ denote the sensitivities of the chiller power to the zonal set-points. In other words, the effect of a change ($\Delta T_{\text{set},z}$) in zonal set-point on the HVAC chiller power consumption is given by $\Delta P_{chiller}\!=\!\beta_z\,\Delta T_{\text{set},z}$. Note that the rationale behind the relation \eqref{E:chiller} is the fact that the heat exchange between two zones separated by a wall is proportional to the difference between the zone temperatures (which in thermal equilibrium is equal to the zone set-point), and the heat exchange for any peripheral zone with the outside is similarly proportional to the difference between the zone temperature and the outdoor air temperature. Collecting time-series data-points on the chiller power, zonal set-points and the outdoor temperature values, we solve a linear regression (least squares) problem to identify the best estimates of the parameters $\beta_0,\,\beta_\text{out}$ and $\beta_z\,\forall z$\,. Fig.\,\ref{fig:chiller} shows the performance of the estimator \eqref{E:chiller} - using both time-series comparisons, as well as estimation error statistics. 
%

\begin{remark}
The simplistic model \eqref{E:chiller} used in this paper is only illustrative, and other (possibly more complex) models can also be integrated into the prioritization framework.
\end{remark}

\begin{figure}[thpb]
\centering
\includegraphics[width=0.9\linewidth]{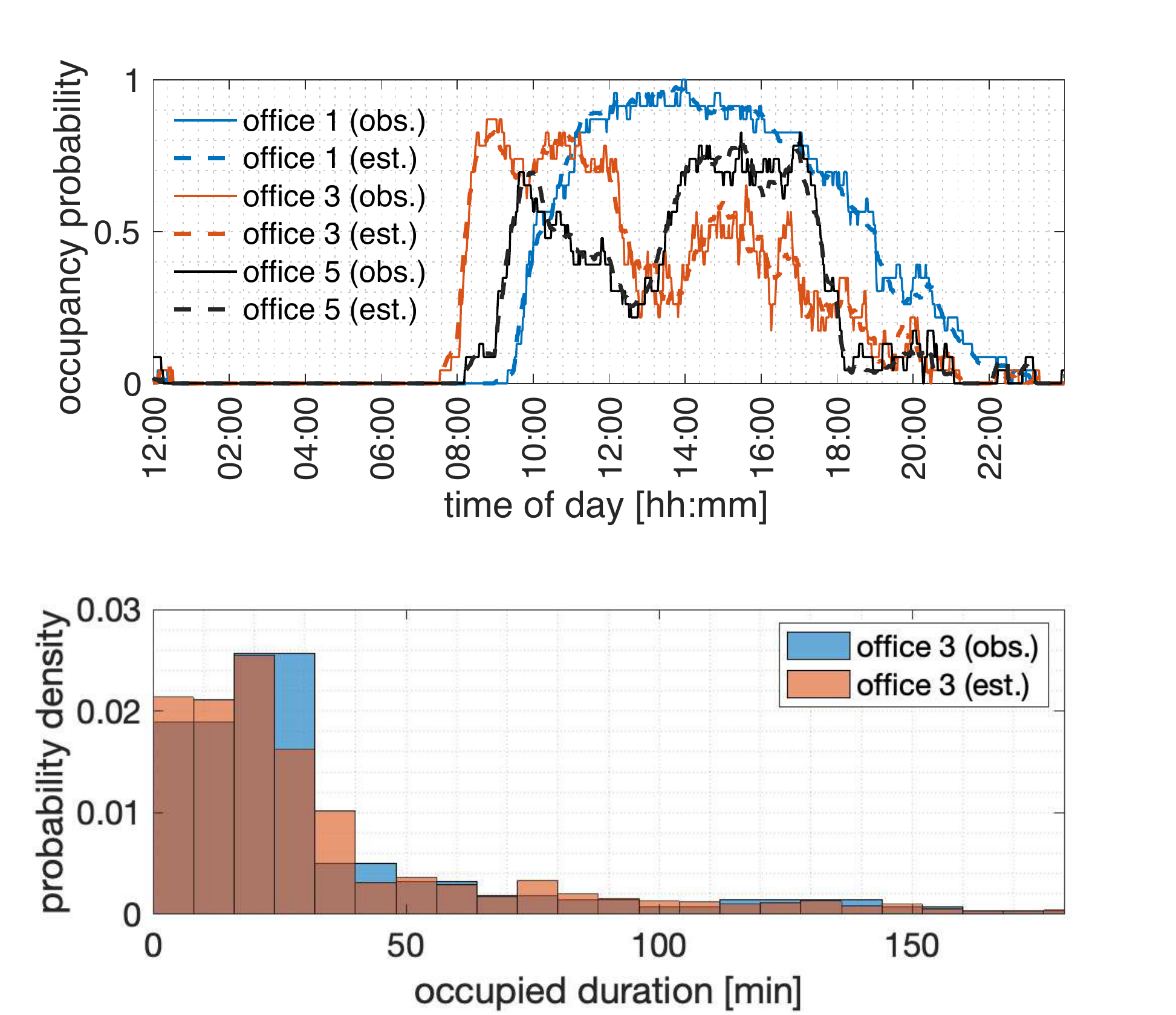}
\caption{(Occupancy modeling) Top plot demonstrates the agreement between the observed and the estimated daily occupancy patterns (likelihood of being occupied) for three different zones. Bottom plot illustrates the closeness between the probability densities of the occupancy duration for one zone.}
\label{fig:occ}
\end{figure}

\textbf{Learning Dynamic Models:} Collection of `historical' and/or experimental data-sets, including time-series sequences of outdoor and zonal temperatures, temperature set-points, power measurements, occupancy, and other contextual information (such as day of week, time of day), are used to develop stochastic (Markov) models for forecasting the impact of various control choices on the criteria scores. Similar Markov-based modeling approach was followed for residential building applications in \cite{vivekanathan2015,kundu2021stochastic} to identify predictive dynamic models for room temperatures, and plug-loads. However, for the multi-zone commercial building application in this paper, we additionally need to learn the occupancy patterns for each zone. Specifically, we adopt the non-homogeneous Markov chain model for occupancy as discussed in \cite{andersen2014dynamic,salimi2019occupancy} (and the references therein). A binary occupancy prediction model is used, wherein the occupancy in each zone is characterized by either `0' (unoccupied) or `1' (occupied). Duration of the occupied and occupied states are accounted for in constructing the states of the Markov model. The day is divided into equal 30-minutes time windows, and a transition probability matrix is learnt for each of the time windows, resulting in a set of 48 Markov models stitched together to form the non-homogeneous Markov chain occupancy model (separately) for each zone. Realistic commercial building occupancy data for six different zones obtained from \cite{dong2015}, event-triggered at fine temporal resolution, are used as historical data for learning the occupancy models. Details are omitted due to space constraint, but Fig.\,\ref{fig:occ} illustrates the performance of the identified occupancy models. Close agreement between the model output and the observed data (from \cite{dong2015}) is seen using both the daily occupancy patterns (i.e. the likelihood of being occupied at each time of a weekday) and the probability distributions of the occupied duration.

%

\subsection{Prioritization Algorithm}\label{sec:algorithm}

For prioritizing the control alternatives based on the predictive models developed in Sec.\,\ref{sec:modeling}, we adopt a stochastic multi-criteria decision-making (MCDM) algorithm from \cite{fan2010,hwang2012}. The method has recently been used for ranking controllable loads in residential households, in the context of both aggregated residential air-conditioning loads \cite{vivekanathan2015} as well as single households with air-conditioner, plug-in electric vehicle and plug-loads \cite{kundu2021stochastic}. For the sake of completeness, we present only a brief description of the algorithm in this paper, while referring the interested readers to \cite{fan2010,hwang2012,vivekanathan2015,kundu2021stochastic} for additional details. Recalling the notations in Sec.\,\ref{sec:problem}, we denote by the (finite and discrete) random variable $X_a^n$ the performance score of control alternative $n\!\in\!\calN$ evaluated on criterion $a\!\in\!\calA$. The criteria scores are mutually independent (for every alternative), since the comfort score relies on occupant perception while curtailment score is related to power ratings. Let $E_a^{n>m}$ be the event that $n$-th alternative outscores the $m$-th one based on the $a$-th criterion. We say that $E_a^{n>m}$ happens with probability 1, 0.5 or 0 when, respectively, the event $X_a^n\!>\!X_a^m,\,X_a^n\!=\!X_a^m$ or $X_a^n\!<\!X_a^m$ occurs. Below is a description of a three-step procedure to compute the ranks.

{\bf{Step 1}} \textsc{(pairwise comparison)}: Note that, by definition, $\pr\left(E_a^{n>m}\right)$ denotes the probability that the $n$-th alternative outscores the $m$-th one with respect to the $a$-th criterion as; and that $\pr\left(E_a^{m>n}\right)\!=\!1-\pr\left(E_a^{n>m}\right)$\,. For every pair of alternatives $n,m\!\in\!\calN$ and for each criterion $a\in\calA$, define the following indicator variable with respect to the event $E_a^{n>m}$\,:
\begin{align*}
    g_a^{n,m} = &\one_{E_a^{n>m}}\,,~\text{ such that }~\pr\left(g_a^{n,m}\!=1\!\right)\!=\!\pr\left(E_a^{n>m}\right)\,.
\end{align*}
Moreover, define $G^{n,m}\!\equiv\!\left(g_a^{n,m}\!: a \!\in\!\calA\right)\!$ as a vector of the indicator variables. Note that $G^{n,m}$ provides an overall summary of the relative superiority of the $n$-th alternative over the $m$-th one, based on all the criteria. There are $2^A$ possible realizations of $G^{n,m}$, with the probability of occurrence of the $h$-th realization given by (for mutually independent criteria):
\begin{align}
    q_h^{n,m}\!=\!\pr\left(G^{n,m}\!=\!G_h^{n,m}\right)\!=\!\prod_{a\in\mathcal{A}}\pr\left(g_a^{n,m}\!=\!g_{a,h}^{n,m}\right).
\end{align}

{\bf{Step 2}} \textsc{(classification of outcomes)}: For every ordered pair $(n,m)$, the possible outcomes of the pairwise comparison vectors classified into three mutually exclusive sets, namely \textit{Most Preferable} ($S_1^{n,m}$), \textit{Indifferent} ($S_2^{n,m})$, and \textit{Not Preferable} ($S_3^{n,m}$), using a linear classifier involving \textit{apriori} criteria weights selected by the decision-maker. Specifically, the $h$-th realization of $G^{n,m}$ is classified as follows:
\begin{align}
    \!\!\forall h\!\in\!\lbrace 1,\dots,2^A\rbrace:~h\!\in\!\left\lbrace \begin{array}{cl}
    S_1^{n,m},        & \, \text{if }W^TG_h^{n,m}\!>\!\nu\\
    S_3^{n,m},        & \, \text{if }W^TG_h^{n,m}\!<\!1\!-\!\nu\\
    S_2^{n,m},        & \, \text{otherwise}
    \end{array}\right.\!\!
\end{align}
where $W \!\equiv\! \left(w_a:a\!\in\!\mathcal{A}\right)$ is a vector of normalized criteria weights (i.e. $\sum_{a\in\calA}w_a=1$), and $\nu\!\in\!(0.5,1)$ is some classification threshold. Further note that,  \[\pr\left(h\!\in\!S_i^{n,m}\right)\!=\!{\sum}_{h\in S_i^{n,m}}\,q_h^{n,m}\,.\]


{\bf{Step 3}} \textsc{(ranking)}: The relative overall superiority of alternative $n$ over $m$, across all criteria, is evaluated as 
\[r^{n>m}= \pr\left(h\!\in\!S_1^{n,m}\right) + 0.5 \,\pr\left(h\!\in\!S_2^{n,m}\right), \quad \forall n\!\neq\!m\,.\]
We define the \textit{fitness values} for each alternative $n$ as follows: 
\begin{align}
    \text{(fitness values)}\quad f^n\!:=\!\frac{1}{N\!-\!1}{\sum}_{m\neq n}r^{n>m}
\end{align}
The fitness value ($f^n$) is an upper bound on the probability of device $n$ being the most superior\footnote{Consider a set of $K$ random variables $\lbrace X_1,\dots,X_K\rbrace$. For any pair $(X_k,X_l)$, we have $\pr(X_k\!>\!\max_{i\neq k}X_i)\!\leq\!\pr\left(X_k\!>\!X_l\right)$. It therefore follows that: $\pr\left(X_k\!>\!\max_{i\neq k}X_i\right)\leq\sum_{i\neq k}\pr\left(X_k\!>\!X_i\right)/(K\!-\!1)$.}, and therefore gives an optimistic estimate on the likelihood of superiority of that device. Finally, the control alternatives are ranked in the descending order of their fitness values, $\lbrace f^n\rbrace_{n=1}^N$..

\begin{remark}[\textbf{Computational Complexity}]
The MCDM algorithm involves algebraic calculations of the pairwise comparison probabilities, for all possible realizations of $G^{n,m}$ (for any pair $n,m$), and therefore has quadratic ($O[N^2]$) complexity with respect to the number of alternatives. Even though the number of possible realizations of $G^{n,m}$ grows exponentially with the number of criteria, it is usually not a concern since the number of criteria is typically fixed and small. As reported in \cite{kundu2021stochastic}, the algorithm takes about 30\,ms to compute ranking when $N\!=\!120$ and the number of criteria is set to three, offering scalability appropriate for commercial building applications.
\end{remark}
\section{Simulation-Based Testing System}
\label{sec:system_architecture}

\begin{figure*}[t]
    \centering
    \begin{minipage}{0.535\textwidth}
    \centering
        \includegraphics[width=0.9\textwidth]{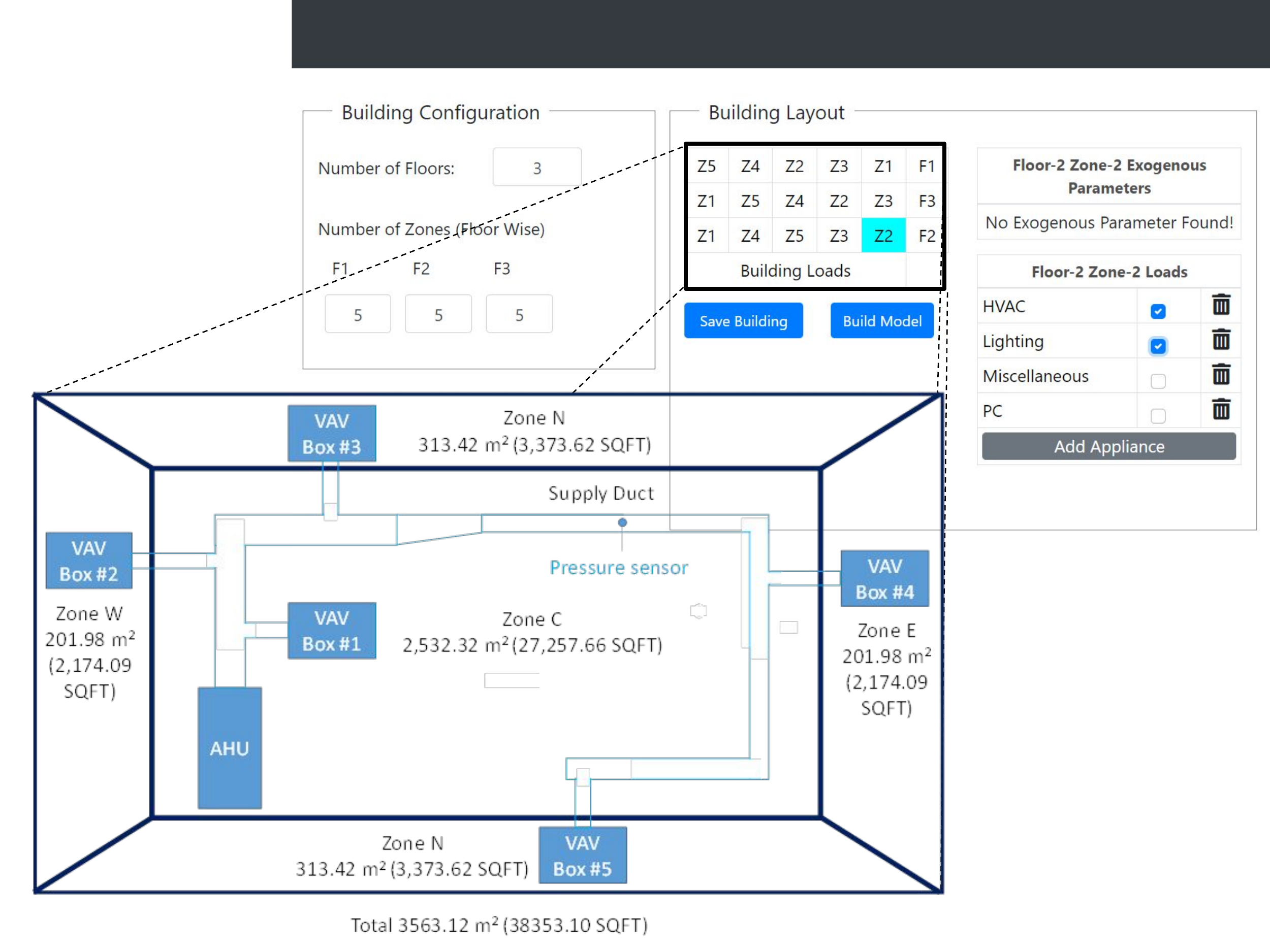}
    \end{minipage}
    \begin{minipage}{0.45\textwidth}
    \centering
        \includegraphics[width=0.8\textwidth]{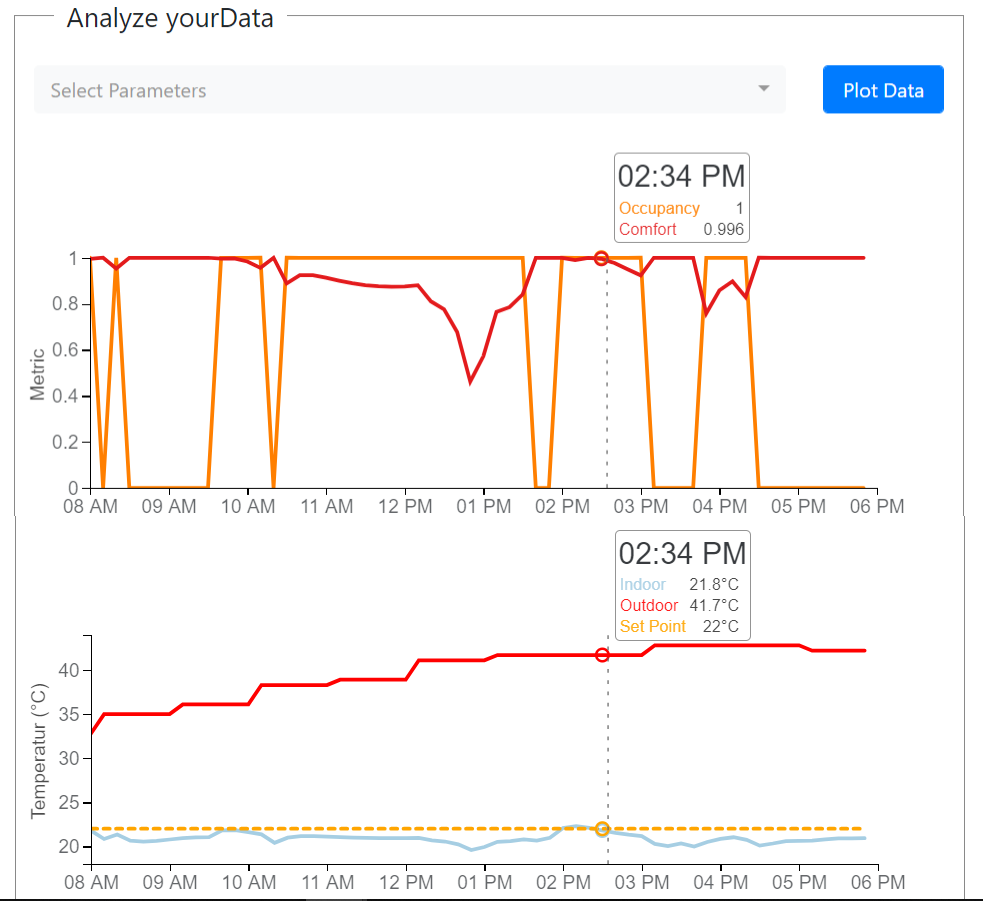}
    \end{minipage}
    \caption{The frontend comprises of two major components - (left) a building configuration screen to review or edit the building, and (right) a visual analytics tool to analyze the data. The frontend plays a key role in ensuring user involvement in the building-grid integration. Not only the frontend allows user to receive the ranking updates, but also provide a medium to analyze the feedback to encourage active engagement and participation.}
    \label{fig:frontend}
\end{figure*}


We implemented a test prototype of the system shown in Fig.\,\ref{fig:system_architecture} that allows the user to interact with a simulated building environment (i.e. a digital twin), analyze the sensor data, visualize ranking recommendations, adjust preferences, and control the prioritized loads to provide grid services. The prototype framework comprises of four major components a building emulator, a controller for analysis and decision automation, a frontend for user interaction, and a backend server to manage the data and control flow between components.

The building \textbf{\textit{emulator}} is designed as a digital twin of a typical large office building. This office building consists of three floors (five zones per each floor) with a total floor area of 46,320$m^2$. The model is implemented in Modelica and can take external control signals and building occupant activity pattern generated at runtime. This emulator provides a high-fidelity representation of the appliance, equipment, and system usage patterns and associated electricity consumption by this building with physics-based models. To facilitate the usage of this emulator for control-related applications, Docker is used to encapsulate the running environment and a REST API is implemented standardizing the control interface. Detailed information about the emulator and the software framework is provided in~\cite{huang2018control}  and~\cite{blum2019prototyping}, respectively.
%
%
The \textbf{\textit{controller}} module executes three functions: a) data-driven modeling: available measurements and contextual information such as occupants' schedule, outside weather conditions are used to develop predictive models of the utilization and associated power consumption of the devices and systems selected for evaluation; b) multi-criteria prioritization: the predictive models are used to estimate the performance scores of the devices across the multiple criteria, informing an MCDM algorithm to rank the devices; c) control computations: the ranking results and simulation data are used to computes control decisions to change the operation of the building equipment and devices to provide grid service. In our prototyping and testing framework the controller module interacts with the emulator via a REST API. The modeling, ranking and control code is written in Julia.
%
The \textbf{\textit{frontend}} enables user interaction for providing building information, configuring the data analytics, ranking preferences, selecting the control objective, and provides performance visualization. 
As shown in Fig.\,\ref{fig:frontend}, users can specify the building configuration, 
add or remove appliances and systems in each zone, and visualize the ranking and control results.  
User interaction is an important aspect of integration of building operation with the smart grid infrastructure, enabling operators and/or owners to apprehend the rationale behind the suggested rankings and make decisions and adopt energy-efficient behaviors. 
The frontend in this prototype system is implemented in Python using Flask. 
%
The \textbf{\textit{backend}} server manages the control and the data flow between other modules of the framework. The backend interacts over \emph{http} requests with the emulator and the frontend, and over \emph{tcp} socket with the controller. The backend is implemented in Python. 
\section{Empirical Evaluation}
\label{sec:results}

To test the efficacy of the ranking method and the use of the ranking results to compute control decisions we considered an energy curtailment scenario. The office building emulator models three floors, each with five zones, for a total of fifteen zones. From each zone, the appliances selected as curtailable are: HVAC (via zone set-point), dimmable lights, and computers (PC). The temperature set-points are allowed to be changed in discrete steps of $\pm1^oC$, not exceeding $\pm5^oC$. Each dimmable light is similarly allowed to be changed (reduced) in discrete steps of $20\%$. The computer (PC) is treated as a plug-load allowing only switch on/off control commands. Only the time window of 8:00AM to 4:00PM is considered for curtailment purposes as that is the time interval when a typical office building is occupied and in use.
%
%
%
\begin{figure}[thpb]
\centering
\vspace{-0.2in}
\subfigure[time-series of building load]{
\includegraphics[width=0.9\linewidth]{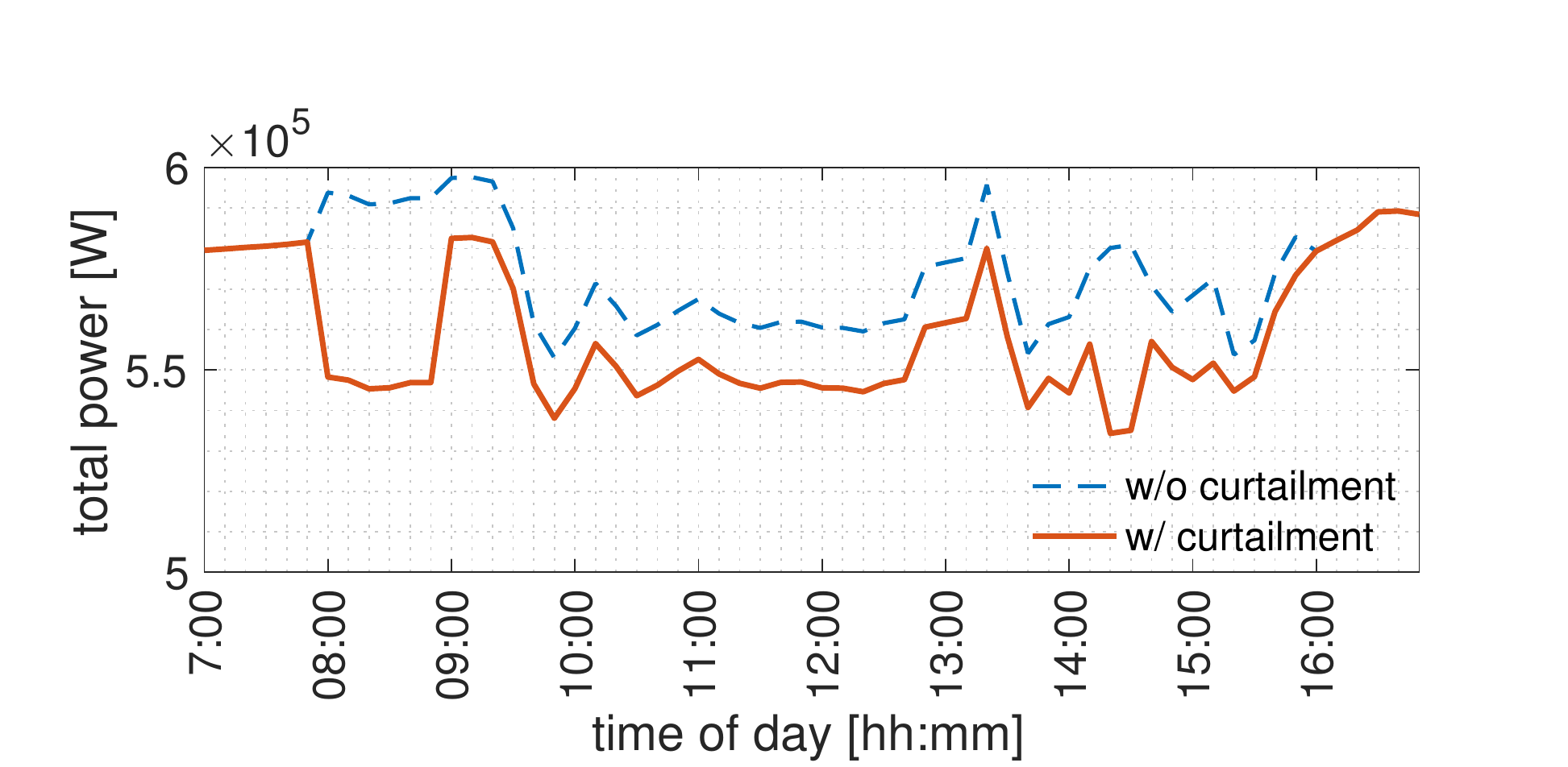}\label{F:power}
}
\subfigure[time-series data of a particular zone]{
\includegraphics[width=0.9\linewidth]{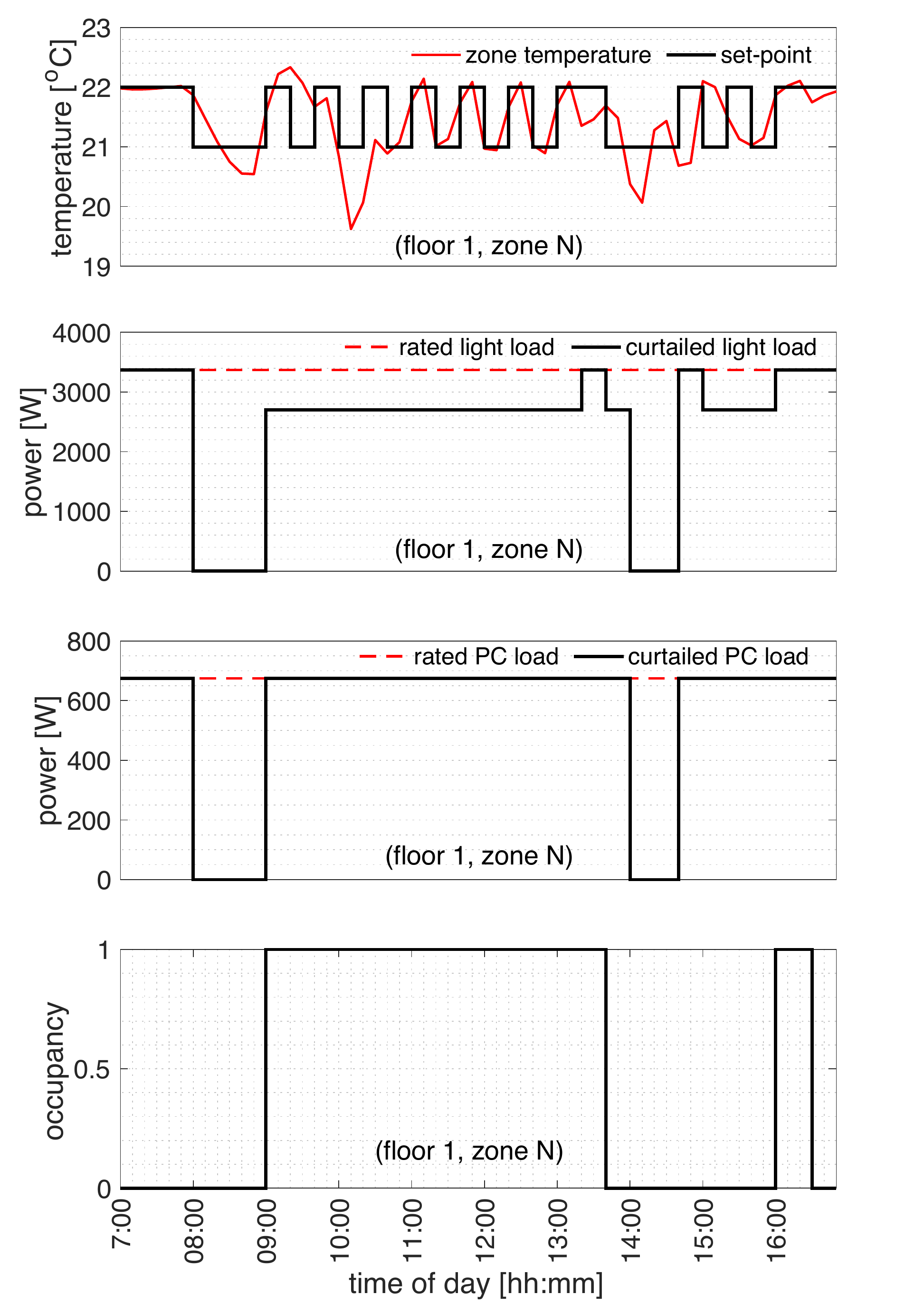}\label{F:zone}
}
\caption[]{Illustrative time-series results from an application of the proposed algorithm to the commercial building use-case. The plot (a) shows the time-series of the total power as a curtailment, in comparison to an estimate of the total power without any curtailment. The plots in (b) display the time-series data related to the curtailable loads: HVAC, lighting and PC, from a particular zone (floor 1, zone N), as well as the occupancy sequence. The criteria weights chosen are: 0.6 (comfort) and 0.4 (curtailment).}
\vspace{-0.2in}
\label{F:res_6_4}
\end{figure}
Fig.\,\ref{F:res_6_4} shows an illustrative example from an application of the proposed algorithm to the commercial building use-case, when the criteria weights are chosen as 0.6 for comfort, and 0.4 for curtailment. It is seen from Fig.\,\ref{F:power}, that early in the day the building is partially unoccupied (also seen from the occupancy patterns in Fig.\,\ref{fig:occ}), which is correspondingly reflected in relatively large amount curtailment, followed by relatively shallow curtailment periods. For better interpretability, we display the results for a particular zone (at floor-1, zone-N). Initially the zone is unoccupied until around 9:00AM. During this time, both the light and the PC are turned off (drawing zero power), while the zonal temperature is set to low. Between 9:00AM to some time after 1:30PM, the zone remains occupied - which is reflected in the PC being turned on during this whole time, while the light is only dimmed by about 20\%. The zonal set-point keeps fluctuating, always keeping the room temperature reasonably close to desired value. After 2:00PM, the zone becomes intermittently occupied, opening up more opportunities for curtailing lighting and PC load.

\section{Conclusion}
\label{sec:conclusion}

In this paper, we consider the emerging concept of a grid-interactive efficient building in which building loads can be controlled to provide grid services. Considering load curtailment as a specific grid service, we develop and demonstrate a data-driven stochastic decision-support framework for dynamic prioritization of heterogeneous control alternatives in a multi-zone commercial building. A stochastic multi-criteria decision-making algorithm (MCDM) is used to simultaneously satisfy multiple (often, competing) operational criteria, e.g. maintaining occupant comfort or maximizing grid service benefit, under uncertainties. In particular, recognizing the critical role occupancy plays in the end-usage (and hence energy consumption) of a commercial building, we integrate Markov-based occupancy models into the prioritization framework. 
Additionally, the proposed framework allows prioritization of complex and heterogeneous control commands (e.g. changes in set-point, dimming of lights). With the help of a prototype system that integrates an interactive \textit{Data Analytics and Visualization} frontend we demonstrate how building operators (and/or owners) can monitor building operations and recommendations, and enable trust-augmented real-time decision via interpreting the rationale behind the ranking. 
Future work will consider a wider set of grid services, extension to smart communities, and testing on real buildings.

\section*{Acknowledgment}
The work was supported by the Building Technologies Office of the U.S. Department of Energy (contract no. DEAC05-76RL01830).


\bibliographystyle{IEEEtran}
\bibliography{main.bib}

\begin{thebibliography}{10}
\providecommand{\url}[1]{#1}
\csname url@samestyle\endcsname
\providecommand{\newblock}{\relax}
\providecommand{\bibinfo}[2]{#2}
\providecommand{\BIBentrySTDinterwordspacing}{\spaceskip=0pt\relax}
\providecommand{\BIBentryALTinterwordstretchfactor}{4}
\providecommand{\BIBentryALTinterwordspacing}{\spaceskip=\fontdimen2\font plus
\BIBentryALTinterwordstretchfactor\fontdimen3\font minus
  \fontdimen4\font\relax}
\providecommand{\BIBforeignlanguage}[2]{{%
\expandafter\ifx\csname l@#1\endcsname\relax
\typeout{** WARNING: IEEEtran.bst: No hyphenation pattern has been}%
\typeout{** loaded for the language `#1'. Using the pattern for}%
\typeout{** the default language instead.}%
\else
\language=\csname l@#1\endcsname
\fi
#2}}
\providecommand{\BIBdecl}{\relax}
\BIBdecl

\bibitem{conti2018annual}
{U.S. Energy Information Administration}, ``{Annual Energy Outlook 2018},''
  {U.S. Department of Energy}, Tech. Rep., 2018.

\bibitem{roth2019grid}
A.~Roth and J.~Reyna, ``Grid-interactive efficient buildings technical report
  series: Whole-building controls, sensors, modeling, and analytics,'' National
  Renewable Energy Lab., CO (United States), Tech. Rep., 2019.

\bibitem{neukomm2019grid}
M.~Neukomm, V.~Nubbe, and R.~Fares, ``Grid-interactive efficient buildings
  technical report series: Overview of research challenges and gaps,'' U.S.
  Department of Energy, Tech. Rep., 2019.

\bibitem{hao2015aggregate}
H.~{Hao}, B.~M. {Sanandaji}, K.~{Poolla}, and T.~L. {Vincent}, ``Aggregate
  flexibility of thermostatically controlled loads,'' \emph{IEEE Trans. on
  Power Systems}, vol.~30, no.~1, pp. 189--198, Jan 2015.

\bibitem{vivekanathan2015}
C.~{Vivekananthan} and Y.~{Mishra}, ``Stochastic ranking method for
  thermostatically controllable appliances to provide regulation services,''
  \emph{IEEE Trans. on Power Systems}, vol.~30, no.~4, pp. 1987--1996, 2015.

\bibitem{espinosa2017aggregate}
L.~A.~D. Espinosa, M.~Almassalkhi, P.~Hines, and J.~Frolik, ``Aggregate
  modeling and coordination of diverse energy resources under packetized energy
  management,'' in \emph{2017 IEEE 56th Annual Conference on Decision and
  Control (CDC)}.\hskip 1em plus 0.5em minus 0.4em\relax IEEE, 2017, pp.
  1394--1400.

\bibitem{nandanoori2018prioritized}
S.~P. Nandanoori, S.~Kundu, D.~Vrabie, K.~Kalsi, and J.~Lian, ``Prioritized
  threshold allocation for distributed frequency response,'' in \emph{2018 IEEE
  Conference on Control Technology and Applications}, 2018, pp. 237--244.

\bibitem{hu2020priority}
X.~Hu and J.~Nutaro, ``{A Priority-Based Control Strategy and Performance Bound
  for Aggregated HVAC-Based Load Shaping},'' \emph{IEEE Trans. on Smart Grid},
  2020.

\bibitem{jin2017}
X.~Jin, K.~Baker, D.~Christensen, and S.~Isley, ``Foresee: {A} user-centric
  home energy management system for energy efficiency and demand response,''
  \emph{Applied Energy}, vol. 205, pp. 1583 -- 1595, 2017.

\bibitem{kundu2021stochastic}
S.~Kundu, A.~Bhattacharya, V.~Chandan, N.~Radhakrishnan, V.~Adetola, and
  D.~Vrabie, ``{A Stochastic Multi-Criteria Decision-Making Algorithm for
  Dynamic Load Prioritization in Grid-Interactive Efficient Buildings},''
  \emph{ASME Letters in Dynamic Systems and Control}, vol.~1, no.~3, Mar 2021.

\bibitem{azar2015aggregated}
A.~G. Azar, R.~H. Jacobsen, and Q.~Zhang, ``Aggregated load scheduling for
  residential multi-class appliances: Peak demand reduction,'' in \emph{12th
  Intl. Conference on the European Energy Market}, 2015, pp. 1--6.

\bibitem{weng2011managing}
T.~Weng, B.~Balaji, S.~Dutta, R.~Gupta, and Y.~Agarwal, ``Managing plug-loads
  for demand response within buildings,'' in \emph{3rd ACM Workshop on Embedded
  Sensing Systems for Energy-Efficiency in Buildings}, ser. BuildSys, 2011, pp.
  13--18.

\bibitem{nutaro2016simulation}
J.~Nutaro, O.~Ozmen, J.~Sanyal, D.~Fugate, and T.~Kuruganti, ``Simulation based
  design and testing of a supervisory controller for reducing peak demand in
  buildings,'' in \emph{High Performance Buildings Conference}, 2016.

\bibitem{kim2016behind}
W.~Kim, S.~Katipamula, R.~G. Lutes, and R.~M. Underhill, ``Behind the meter
  grid services: {Intelligent Load Control},'' Pacific Northwest National Lab.,
  Tech. Rep., 2016.

\bibitem{rea2000iesna}
M.~S. Rea \emph{et~al.}, ``{The IESNA Lighting Handbook: Reference \&
  Application},'' 2000.

\bibitem{andersen2014dynamic}
P.~D. Andersen, A.~Iversen, H.~Madsen, and C.~Rode, ``Dynamic modeling of
  presence of occupants using inhomogeneous markov chains,'' \emph{Energy and
  buildings}, vol.~69, pp. 213--223, 2014.

\bibitem{salimi2019occupancy}
S.~Salimi, Z.~Liu, and A.~Hammad, ``Occupancy prediction model for open-plan
  offices using real-time location system and inhomogeneous markov chain,''
  \emph{Building and Environment}, vol. 152, pp. 1--16, 2019.

\bibitem{dong2015}
B.~Dong, ``{Long-term occupancy data for residential and commercial
  building},''
  \url{https://openei.org/datasets/dataset/long-term-occupancy-data-for-residential-and-commercial-building},
  {Accessed: 2019-08-21}.

\bibitem{fan2010}
Z.-P. Fan, Y.~Liu, and B.~Feng, ``A method for stochastic multiple criteria
  decision making based on pairwise comparisons of alternatives with random
  evaluations,'' \emph{European J. of Operational Research}, vol. 207, pp.
  906--915, 2010.

\bibitem{hwang2012}
C.-L. Hwang and K.~Yoon, \emph{Multiple attribute decision making: {M}ethods
  and applications}.\hskip 1em plus 0.5em minus 0.4em\relax Springer Verlag,
  New York, 2012, vol. 186.

\bibitem{huang2018control}
S.~Huang, Y.~Chen, P.~W. Ehrlich, and D.~L. Vrabie, ``{A control-oriented
  building envelope and HVAC system simulation model for a typical large office
  building},'' in \emph{Proceedings of Building Performance Modeling Conference
  and SimBuild}, Chicago, IL, Sep. 2018.

\bibitem{blum2019prototyping}
D.~Blum, F.~Jorissen, S.~Huang \emph{et~al.}, ``{Prototyping the BOPTEST
  framework for simulation-based testing of advanced control strategies in
  buildings},'' in \emph{Int. Building Performance Simulation Assoc. Conf.},
  2019.

\end{thebibliography}


\end{document}